\documentclass[prb,aps,twocolumn,superscriptaddress,longbibliography]{revtex4-2}

\usepackage[T1]{fontenc}
\usepackage[utf8]{inputenc}
\usepackage{amssymb}
\usepackage{graphicx}
\usepackage{amsmath,color}
\usepackage{mathrsfs}
\usepackage{float}
\usepackage{indentfirst}
\usepackage{subfigure}
\usepackage{multirow}
\usepackage{tabu}
\usepackage{threeparttable}
\usepackage{booktabs}
\usepackage{txfonts}
\usepackage{amssymb,bbm}
\usepackage[normalem]{ulem}
\usepackage[euler]{textgreek}
\usepackage{bm}% bold math

\usepackage[dvipsnames]{xcolor}
\usepackage{soul}
\definecolor{linkcolour}{rgb}{0.0,0.2,0.6}
\usepackage[colorlinks=true,urlcolor=linkcolour,citecolor=linkcolour,linkcolor=linkcolour]{hyperref}

\newcommand{\mathd}{\mathrm{d}}

\newcommand{\mathi}{\mathrm{i}}

\newcommand{\calP}{\mathcal{P}}

\newcommand{\barQ}{\bar{Q}}
\newcommand{\barR}{\bar{R}}

\newcommand{\hatH}{\hat{H}}

\newcommand{\id}{\mathbbm{1}}

\newcommand{\tr}{\mathrm{Tr}}

\newcommand\redsout{\bgroup\markoverwith{\textcolor{red}{\rule[0.5ex]{2pt}{0.4pt}}}\ULon}

\newcommand{\ket}[1]{\left|#1\right\rangle}
\newcommand{\bra}[1]{\left\langle#1\right|}

\begin{document}

\preprint{APS/123-QED}

\title{Numerical study of boson mixtures with multi-component continuous matrix product states}

\author{Wei Tang}
\email{wei.tang.phys@gmail.com}
\affiliation{Department of Physics and Astronomy, Ghent University, Krijgslaan 281, 9000 Gent, Belgium}

\author{Beno\^{i}t Tuybens}
\affiliation{Department of Physics and Astronomy, Ghent University, Krijgslaan 281, 9000 Gent, Belgium}

\author{Jutho Haegeman}
\affiliation{Department of Physics and Astronomy, Ghent University, Krijgslaan 281, 9000 Gent, Belgium}

\date{\today}

\begin{abstract}
The continuous matrix product state (cMPS) ansatz is a promising numerical tool for studying quantum many-body systems in continuous space. 
Although it provides a clean framework that allows one to directly simulate continuous systems, the optimization of cMPS is known to be a very challenging task, especially in the case of multi-component systems. 
In this work, we have developed an improved optimization scheme for multi-component cMPS that enables simulations of bosonic quantum mixtures with substantially larger bond dimensions than previous works.
We benchmark our method on the two-component Lieb-Liniger model, obtaining numerical results that agree well with analytical predictions. 
Our work paves the way for further numerical studies of quantum mixture systems using the cMPS ansatz.
\end{abstract} 

\maketitle

\section{Introduction} 

The past decades have witnessed remarkable progress in cold atom physics, which has enabled clean and highly controllable experimental realizations of a wide range of quantum many-body systems~\cite{bloch-manybody-2008,daley-practical-2022}.
In optical lattice and engineered atomic lattice setups, ultracold atom experiments constitute ideal platforms for the quantum simulation of a wide range of lattice many-body systems~\cite{bloch-quantum-2012,schafer-tools-2020,browaeys-many-body-2020}.
Alternatively, ultracold atomic gases in free space realize intrinsically continuous quantum systems with tunable interactions, dimensionality, and external confining potentials~\cite{cazalilla-one-2011,daley-practical-2022}.
Furthermore, multi-component quantum mixtures introduce additional degrees of freedom, providing platforms for exploring rich physical phenomena arising from the interplay between inter-species interactions, quantum statistics, and mass imbalance effects~\cite{baroni-quantum-2024}.

The classical simulation of such interacting free-space systems can pose significant challenges, as many numerical approaches require some form of discretization that can introduce spurious effects or be ill-suited for capturing dynamical phenomena, and in any case requires a careful extrapolation to the continuum limit. 
For quasi-one-dimensional gases, the continuous matrix product state (cMPS) ansatz, obtained as a natural continuum limit of the matrix product state (MPS) ansatz, provides a promising numerical framework~\cite{verstraete-continuous-2010}, namely as a class of variational wave functions that are intrinsically defined in continuous space.
As in the case of MPS, the variational parameters in a cMPS correspond to matrices, the size of which is known as the bond dimension and regulates the entanglement in the system, thereby acting as a natural refinement parameter that controls the expressivity of the ansatz. The cMPS method has been applied to study atomic gases in a variety of settings~\cite{verstraete-continuous-2010,draxler-particles-2013,rincon-liebliniger-2015,chung-matrix-2015,ganahl-continuous-2017-a,tuybens-variational-2022,lukin-continuous-2022,tang-kac-2023}, but was also employed to obtain thermodynamic properties of quantum lattice systems~\cite{tang-continuous-2020,tang-tensor-2021,zhang-universal-2023}. Furthermore, (1+1)-dimensional relativistic quantum theories have been studied by either introducing a cutoff~\cite{haegeman-applying-2010}, or by extending the cMPS ansatz using the vacuum and creation operators of the non-interacting part of the theory and regularizing the interactions via normal ordering~\cite{tilloy-relativistic-2021,tilloy-study-2022-arxiv,tiwana-relativistic-2025}.

Importantly, the cMPS ansatz is not limited to single-component systems, but is also applicable to multi-component systems, including boson-boson mixtures, boson-fermion mixtures, and fermion-fermion mixtures.
However, it was shown that, unlike in the single-boson case, in order to obtain sufficiently regular wave functions with finite kinetic energy for such multi-component systems, the matrices that constitute the variational parameters of the cMPS ansatz need to satisfy additional constraints~\cite{haegeman-calculus-2013}. 
While several numerical studies have been performed on quantum mixtures using the multi-component cMPS ansatz~\cite{quijandria-continuous-2014,quijandria-continuous-2015,peacock-quantum-2022}, they are typically limited to relatively small bond dimensions, making it difficult to achieve higher accuracy.
The main obstacle in these multi-component cMPS simulations is the difficulty of optimizing the cMPS parameters. 
Although it might appear that the variational optimization of cMPS is straightforward, the energy landscape is typically very ill-conditioned, making the optimization very slow. 
The additional constraints on the cMPS parameters in the multi-component cMPS make the optimization even more challenging, as they restrict the ability to convert the cMPS into canonical forms.  
Standard gradient-based optimization schemes typically require a huge number of iterations to converge, which quickly becomes prohibitively expensive for large bond dimensions.

On the other hand, it has recently been shown that the variational optimization of tensor network states can be considerably accelerated by incorporating the geometric structure of the tensor network manifold~\cite{hauru-riemannian-2021}.
More specifically, one can use the metric of the tensor network state manifold, \textit{i.e.} the induced Hilbert space inner product on the manifold's tangent space, as a preconditioner.
This can significantly reduce the number of iterations required to converge. 
This preconditioning scheme can be flexibly incorporated into standard gradient-based optimization algorithms, and has been successfully applied in several numerical works~\cite{hauru-riemannian-2021,tang-kac-2023,zhang-accelerating-2025-arxiv}.
In the context of multi-component cMPS, however, it remains unclear how to efficiently construct the corresponding tangent-space metric preconditioner, due to the complexities introduced by the additional constraints imposed on the cMPS parameters, which are not present in single-boson cMPS.

In this work, we propose a new optimization scheme for bosonic multi-component cMPS. 
We introduce a new parametrization for the cMPS matrices such that the regularity condition is automatically satisfied, while the gauge degrees of freedom are still fully preserved.
This allows us to exploit the gauge degrees of freedom and perform the optimization in the left-canonical form, which enables a convenient construction of the metric preconditioner.
Our optimization scheme enables us to perform simulations with substantially larger bond dimensions than previous works, giving rise to more accurate numerical results. 
We benchmark the cMPS method in the two-component Lieb-Liniger model, and compare the cMPS results with analytical predictions obtained from different approximations. 

The remainder of this paper is organized as follows. 
Section \ref{sec:cMPS} provides a brief review of the (bosonic) cMPS ansatz, focusing on its application to systems with multiple boson species. 
In Sec.~\ref{sec:improved-optimization-scheme}, we propose our novel optimization scheme for the multi-component cMPS.
Sec.~\ref{sec:benchmarks} presents a detailed benchmark of the cMPS method in the two-component Lieb-Liniger model, where we compare the cMPS results with analytical predictions obtained with different approximations. 
Finally, Sec.~\ref{sec:summary-and-outlook} provides a summary and outlook.
We provide some additional technical details in the appendices.
In Appendix \ref{sec:penalty-method}, we discuss an alternative optimization strategy via penalty terms. 
Appendix \ref{sec:regularization-parameter-in-preconditioner} addresses the choice of the regularization parameter in the preconditioner.
In Appendix \ref{sec:preparation-of-initial-state}, we discuss the steps for the preparation of initial states. 
Appendix \ref{sec:stiffness} provides details on the computation of the stiffness in the two-component Lieb-Liniger model.

\section{Continuous matrix product states: a brief review}\label{sec:cMPS}

This section provides a brief review of the cMPS formalism for systems with multiple bosonic species.
A continuous matrix product state (cMPS) for multiple bosons defined on the infinite one-dimensional continuous space is a state of the form
\begin{equation}
    \ket{\Psi} = \tr_{\mathrm{aux}} \left[\mathcal{P} \exp \left(\int_{-\infty}^\infty \mathd x \left(Q \otimes \mathbbm{1} + \sum_j R_j \otimes \psi_j^\dagger(x) \right)\right) \right] \ket{\Omega}. \label{eq:multi-boson-cMPS}
\end{equation}
Here, $\psi_j(x)$ is the bosonic field operator for the $j$th bosonic field, $\ket{\Omega}$ is the vacuum state satisfying $\psi_j(x) \ket{\Omega} = 0$ for all $j$ species, and $\mathcal{P}$ represents a path-ordered exponential. 
The bosonic field operators satisfy the commutation relations ($\forall j,k$)
\begin{align}
    \left[\psi_j(x), \psi_k(y)\right] &= 0, \\
    \left[\psi_j(x), \psi_k^\dagger(y)\right] &= \delta_{jk} \delta(x-y).
\end{align}
The matrices $Q$ and $R_j$ are $\chi\times\chi$-dimensional matrices living in the auxiliary space, where $\chi$ is the bond dimension, and a trace is taken over this auxiliary space.
In this work, we have restricted ourselves to infinite homogeneous systems, i.e., the matrices $Q$ and $R_j$ are independent of the position $x$.
Simulations with spatially varying $Q$ and $R_j$ can also be performed, but the computation of physical observables and the optimization of these cMPS states are more involved~\cite{tuybens-variational-2022} and will not be discussed in this work.

In this translation invariant setting, the cMPS parametrization has a gauge invariance, in the sense that the same physical state is obtained when simultaneously changing $Q \to G Q G^{-1}$ and $R_j \to G R_j G^{-1}$ for some invertible matrix $G \in \mathsf{GL}(\chi)$. This gauge freedom can be used to impose a relation between the $Q$ and $R_j$ matrices, such as the \emph{left-canonical form}
\begin{equation}
    Q + Q^\dagger + \sum_j R_j^\dagger R_j = 0. \label{eq:left-canonical-condition}
\end{equation}
This condition states that the $(\chi^2 \times \chi^2)$-dimensional cMPS transfer matrix $T = Q \otimes \id + \id \otimes \overline{Q} + \sum_{j} R_j \otimes \overline{R_j}$ has an exact zero eigenvalue, which will be the eigenvalue with the largest real part, and that the corresponding left eigenvector $\bra{l}$ will be given by $l \sim \id$ when reinterpreted as a $\chi \times \chi$ matrix. Exploiting this canonical form can simplify and stabilize the numerical implementation, and will be an important consideration in the remainder of this manuscript.

When the number of bosonic species is larger than one, the $R_j$ matrices cannot be chosen completely arbitrarily, as this would result in a wavefunction that exhibits discontinuities when positions of different particles approach each other.
This manifests itself in a diverging kinetic energy density.
More specifically, the kinetic energy density of the state defined by Eq.~\eqref{eq:multi-boson-cMPS} is given by
\begin{align}
    \sum_j \left\langle \partial_x \hat{\psi}_j^\dagger(x) \partial_x \hat{\psi}_j(x) \right\rangle_{\Psi} &=
    \sum_j \langle l | \left[\barQ, \barR_j\right] \otimes \left[\barQ, \barR_j\right] | r \rangle  \nonumber \\
    & \hspace{-0.25em} + \delta(0) \sum_{j,k} \langle l | \left[\barR_j, \barR_k\right] \otimes \left[R_j, R_k\right] | r \rangle
    \label{eq:kinetic-energy-divergence}
\end{align}
where $\ket{l}$ and $\ket{r}$ denote the left and right leading eigenvectors of the cMPS transfer matrix, respectively~\cite{haegeman-calculus-2013}.
The second term of Eq.~\eqref{eq:kinetic-energy-divergence} contains a divergent contribution $\delta(0)$. 
As the leading eigenvectors of the transfer matrix correspond to positive definite matrices, this divergence can only be avoided by imposing the following regularity condition on the $R_j$ matrices~\cite{haegeman-calculus-2013}
\begin{equation}
    \left[R_j, R_k\right] = 0. \label{eq:regularity-condition}
\end{equation}
This condition originates from the fact that a cMPS (and MPS alike) can be thought of as generating the physical state sequentially~\cite{PhysRevLett.95.110503,PhysRevLett.105.260401}, as expressed clearly by the path-ordered exponential appearing in the cMPS definition.
While we expect the limit $x \to y$ of $\psi_j(x) \psi_k(y) \ket{\Psi}$ to be well-defined and independent of the order of $x$ and $y$, this does not hold for a cMPS unless Eq.~\eqref{eq:regularity-condition} is satisfied.

The most straightforward approach to impose the regularity condition during a variational ground state search is to find explicit parametrizations of the $R_j$ matrices that satisfy Eq.~\eqref{eq:regularity-condition}.
In previous studies, the $R_j$ matrices have been parametrized in the tensor product form~\cite{quijandria-continuous-2014,quijandria-continuous-2015}
\begin{equation}
    R_j = I_1 \otimes \cdots \otimes I_{j-1} \otimes B_j \otimes I_{j+1} \otimes \cdots \otimes I_N,
    \label{eq:tensor-product-parametrization}
\end{equation}
where $B_j$ is a $\chi_j \times \chi_j$ matrix. 
The total bond dimension is then $\chi = \prod_j \chi_j$.
This tensor product parametrization has a clear physical motivation: when the different boson species do not interact, the ground state can be modelled as a tensor product of single-boson cMPSs, which is equivalent to a multi-component cMPS with $R_j$ matrices in the form of Eq.~\eqref{eq:tensor-product-parametrization}.

Despite this physical intuition, this parametrization has several drawbacks in practice.
First, it does not cover all possible commuting $R_j$ matrices, which significantly restricts the expressivity of the cMPS ansatz.
This can easily be understood by the fact that this parametrization imposes a very specific structure on the eigenvalues of the $R_j$ matrices, where every eigenvalue of $R_j$ appears at least $\prod_{k \neq j} \chi_k$ times. 
In particular, when the coupling between components is strong, it is no longer clear whether the parametrization \eqref{eq:tensor-product-parametrization} is still a reasonable choice.
Second, for a given bond dimension $\chi$, one needs to choose a proper $\chi_j$ for each $j$, and the optimal choice is not clear prior to the actual calculation. 
Thirdly, this parametrization does not transform covariantly under a general gauge transformation with $G \in \mathsf{GL}(\chi)$. Instead, the gauge freedom is reduced down to a tensor product form $G = G_1 \otimes G_2 \otimes \cdots \otimes G_N$, which is not sufficient to impose the left-canonical form.
As a consequence, working directly with the parametrization \eqref{eq:tensor-product-parametrization} in the optimization usually leads to a very slow optimization process, due to the ill-conditioned nature of the energy landscape.
As elaborated upon below, a natural strategy to improve the optimization efficiency is to use the tangent-space metric as a preconditioner.
For the parametrization \eqref{eq:tensor-product-parametrization}, however, constructing this metric explicitly is technically involved and computationally expensive due to the lack of the canonical form, resulting in a significant overhead that largely offsets the benefits of preconditioning.

An alternative approach is to avoid parametrizing the $R_j$ matrices themselves, and instead to regularize the second term in Eq.~\eqref{eq:kinetic-energy-divergence} with a finite cutoff $\Lambda$.
Then, the correct results should be obtained by extrapolating to the $\Lambda \rightarrow \infty$ limit.
The benefit of this approach is that one can avoid the complexities originating from parametrizing $R_j$ matrices, making its implementation as simple as the single-boson cMPS. 
However, the drawback of this approach is that, when $\Lambda$ is large, the regularization term makes the energy landscape extremely ill-conditioned, making the optimization very slow. 
The aforementioned preconditioning scheme cannot adequately address this issue, as it only takes into account the geometric structure of the cMPS ansatz itself, and does not take into account the property of the Hamiltonian.
A more fundamental limitation of the penalty method is that it compromises the strict variational character of the calculation. At any finite value of $\Lambda$, the optimization targets the ground state of a \emph{modified} Hamiltonian rather than that of the original physical model. Consequently, the resulting energy is not guaranteed to provide an upper bound on the true ground-state energy of the target Hamiltonian. 
We will examine this strategy in more detail in Appendix \ref{sec:penalty-method}.

In this work, we continue with the strategy of parametrizing $R_j$ matrices, but with a new parametrization that enables us to mitigate or avoid the aforementioned issues.
The details of this new parametrization are discussed in the next section.

\section{Improved optimization scheme for the multi-boson cMPS ansatz}\label{sec:improved-optimization-scheme}

In this work, we consider the following parametrization of the $R$ matrices
\begin{equation}
    R_j = M D_j M^{-1}, \label{eq:diag_parametrization}
\end{equation}
where $D_j$ is a diagonal $\chi \times \chi$ matrix and $M$ is an invertible $\chi \times \chi$ matrix.
Indeed, if we assume that all the $R$ matrices are diagonalizable, then the regularity condition \eqref{eq:regularity-condition} implies that all the $R_j$'s can be simultaneously diagonalized.
Put differently, this parametrization covers all possible diagonalizable commuting $R$ matrices.
While we could use the cMPS gauge freedom to eliminate $M$ and directly parametrize the $R_j$ matrices as diagonal matrices, we instead prefer to keep $M$ so as to have the ability to impose the left-canonical form.
Working in this gauge is crucial for constructing the tangent-space metric and its associated preconditioner in a simple and numerically stable manner.

In order to preserve the left-canonical form during the optimization, we will use the framework of Riemannian gradient-based optimization methods. In the following, we will discuss the key ingredients of the resulting optimization scheme in detail.

\subsection{Tangent vectors}

In the left-canonical form, the matrices $Q$ and $R_j$ satisfy Eq.~\eqref{eq:left-canonical-condition}. At each point $(Q, \{R_j\})$, one can attach a tangent vector $(V, \{W_j\})$, which should then satisfy the following condition 
\begin{equation}
    V = \mathi K - \sum_j R_j^\dagger W_j, \label{eq:tangent-vector-left-canonical}
\end{equation}
with $K$ a Hermitian matrix, in order to preserve the left-canonical form at first order. However, the Hermitian matrix $K$ can be further removed by also exploiting the subgroup of unitary gauge freedom in the cMPS (which is not fixed by the left-canonical condition).

Since we have parametrized $R_j$ as $R_j = M D_j M^{-1}$, one can derive the expression for $W_j$ by differentiating $R_j$, which yields $\dot{R}_j = M ([(M^{-1} \dot{M}), D_j] + \dot{D}_j ) M^{-1}$. 
Accordingly, each tangent vector $(V, \{W_j\})$ can be parametrized by a set of diagonal matrices $\delta D_j$ and a purely off-diagonal matrix $\delta X$ (zero diagonal entries), such that 
\begin{equation}
    W_j = M \left(\left[\delta X, D_j\right] + \delta D_j \right) M^{-1}, \label{eq:W_j_parametrization}
\end{equation}
and $V$ is given by Eq.~\eqref{eq:tangent-vector-left-canonical} with $K = 0$.
Following the parametrization \eqref{eq:W_j_parametrization}, we can define the inner product between tangent vectors as the Euclidean inner product between the matrices $\delta X$ and $\delta D_j$'s, i.e.,
\begin{align}
    \Bigg( \left(\delta X^a, \left\{\delta D_j^a\right\}\right) ,& \left(\delta X^b, \left\{\delta D_j^b\right\}\right) \Bigg)  = \,  \nonumber \\&\tr\left[\left(\delta X^a\right)^\dagger \delta X^b\right]  + \sum_j \tr\left[\left(\delta D_j^a\right)^\dagger \delta D_j^b\right]. \label{eq:tangent-vector-inner-product}
\end{align}

In the context of gradient-based optimization, a crucial role is played by the (dual) vector of partial derivatives of the energy function with respect to the variational parameters, which we denote as $(\partial Q, \partial M, \{\partial D_j\})$.
We need to ``project'' this vector into the tangent space, \textit{i.e.}\ we associate an actual tangent vector to these partial derivatives by matching the first order expansion of the energy with the inner product definition \eqref{eq:tangent-vector-inner-product}.
This tangent vector is what we refer to as the gradient.
More explicitly, one needs to ensure that the inner product between a random tangent vector $(\delta X^r, \{\delta D_j^r\})$ and the gradient $(\delta X^g, \{\delta D_j^g\})$ is equal to the infinitesimal change of the energy function along the direction defined by $(\delta X^r, \{\delta D_j^r\})$, which can be computed using the partial derivatives $(\partial Q, \partial M, \{\partial D_j\})$ of the energy function.
After some arithmetic, we obtain the expression for the gradient 
\begin{align}
    \delta X^g &= M^\dagger \sum_j \left( R_j^\dagger R_j \partial Q - R_j \partial Q R_j^\dagger \right) \left(M^{-1}\right)^\dagger + M^\dagger \partial M, \\
    \delta D_j^g &= -M^\dagger R_j \partial Q \left(M^{-1}\right)^\dagger +  \partial D_j . 
\end{align}

\subsection{Retraction}

Given a search direction in the form of a tangent vector $(V, \{W_j\})$ parametrized by $(\delta X, \{\delta D_j\})$, we need to be able to update the cMPS parameters $(Q, R_j = M D_j M^{-1})$ along this direction with a variable step size $\alpha$, thereby preserving the left-canonical form. The resulting path $(Q(\alpha), R_j(\alpha))$ on the manifold is known as a retraction. The formalism of Riemannian optimization does not impose specific requirements on this retraction, other than that the derivative $(\dot{Q}(0), \dot{R}_j(0))$ matches with the given search direction.

We choose the retraction such that $M$ and $D_j$ are updated as
\begin{align}
    M(\alpha) &= M \exp(\alpha \delta X), \label{eq:retraction-M} \\
    D_j(\alpha) &= D_j + \alpha \delta D_j. \label{eq:retraction-D}
\end{align}
The specific choice for $M(\alpha)$ is such that $M(\alpha)$ cannot acquire accidental zero eigenvalues, which would make it singular. 

We then obtain $R_j(\alpha) = M(\alpha) D_j(\alpha) M(\alpha)^{-1}$, and further update $Q$ so as to preserve the left-canonical condition in Eq.~\eqref{eq:left-canonical-condition} as
\begin{equation}
    Q(\alpha) = Q - \sum_j \left( R^\dagger_j \Delta R_j(\alpha) + \frac{1}{2} \Delta R_j (\alpha)^\dagger \Delta R_j (\alpha) \right), \label{eq:retraction-Q}
\end{equation}
where 
$\Delta R_j (\alpha) = R_j(\alpha) - R_j$.

\subsection{Vector transport}

If one makes use of Riemannian quasi-Newton methods for the optimization, it is also necessary to be able to transport tangent vectors from previous optimization steps to the current optimization step, which is referred to as the vector transport. 
In our implementation, we choose the vector transport to simply act as the identity transformation on the chosen tangent vector parametrization, \textit{i.e.}\ $(\delta X, \{\delta D_j\}) \rightarrow (\delta X, \{\delta D_j\})$.

\subsection{Preconditioner}

Recall that a tangent vector $(\delta X, \{\delta D_j\})$ corresponds to a state $|\Phi(Q, \{R_j\}; \delta X, \{\delta D_j\})\rangle$, given by~\cite{haegeman-calculus-2013,draxler-particles-2013}
\begin{align}
\left|\Phi\left(Q, \{R_j\}; \delta X, \{\delta D_j\}\right)\right\rangle  = \int_{-\infty}^\infty \mathd x \tr_{\mathrm{aux}} \left[ \hat{U}(-\infty, x) \right. \hspace{2em} \nonumber \\
     \left. \times \left(V \otimes \mathbbm{1} + \sum_j W_j \otimes \psi_j^\dagger(x)\right) \hat{U}(x, \infty)\right] \ket{\Omega},
\end{align}
where $\hat{U}(a, b) = \calP \exp [ \int_a^b \mathd x (Q\otimes \mathbbm{1} + \sum_j R_j \otimes \psi_j^\dagger(x)) ]$, and $V$ and $W_j$ are given by Eq.~\eqref{eq:tangent-vector-left-canonical} and Eq.~\eqref{eq:W_j_parametrization}, respectively.
Note that we have employed a rather unphysical definition for the tangent vector overlaps in Eq.~\eqref{eq:tangent-vector-inner-product}. 
This definition is convenient to work with in practice, but it completely ignores the geometric structure of the cMPS manifold. 
To compensate for this, we use the metric $\mathcal{N}$ defined from the physical overlap between the tangent-vector states as a preconditioner. 
When the cMPS is in the left-canonical form and the tangent vectors are parametrized such that $V=-\sum_{j} R_j^\dagger W_j$, the metric $\mathcal{N}$ takes a relatively simple form, which is given by
\begin{align}
    \mathcal{N}_{a, b} &= \left\langle \Phi\left(Q, \{R_j\}; \delta X^a, \{\delta D_j^a\}\right) \middle| \Phi\left( Q, \{R_j\}; \delta X^b, \{\delta D_j^b\}\right) \right\rangle  \nonumber \\
    &= \sum_{j} \tr\left[ \left(W^a_j\right)^\dagger W^b_j \rho_R\right] \label{eq:preconditioner-tangent-vectors-left-canonical},
\end{align}
where $\rho_R$ is the right environment of the cMPS, and $W^a_j$ and $W^b_j$ are given by Eq.~\eqref{eq:W_j_parametrization}.  
Once the metric preconditioner $\mathcal{N}$ is obtained, one can apply its inverse to the gradient to get the preconditioned gradient.

During the optimization, it is often beneficial to regularize the metric preconditioner $\mathcal{N}$ by adding a small regularization term to it, 
i.e., $\mathcal{N} \rightarrow \mathcal{N} + \lambda \mathcal{I}$, where $\mathcal{I}$ is the identity matrix, and $\lambda$ is a small regularization parameter~\cite{hauru-riemannian-2021,tang-kac-2023,zhang-accelerating-2025-arxiv}.
We discuss the choice of $\lambda$ in more detail in Appendix \ref{sec:regularization-parameter-in-preconditioner}.

While the metric looks particularly simple in Eq.~\eqref{eq:preconditioner-tangent-vectors-left-canonical} and is fully specified by the $\chi \times \chi$ matrix $\rho_R$, it ultimately needs to be formulated and inverted for the actual tangent vector parametrization $(\delta X, \delta D_j)$ of length $\chi (\chi + N-1)$ that takes the regularity conditions \eqref{eq:regularity-condition} into account, where $N$ is the number of boson species.
This cannot be avoided and complicates the preconditioning step as compared to the case of the single-boson cMPS.
The full construction and inversion of $\mathcal{N}$ in this parametrization has a computational cost of the order of $O(\chi^6)$, which is higher than the cost $O(\chi^3)$ of each optimization step in ordinary cMPS calculations.
One may consider using an iterative solver to solve the inverse problem, but due to the large condition number in the preconditioner, the convergence of the iterative solver is slow.
Along this line, one can choose not to push the iterative solver to full convergence, but make use of the non-converged solution as an approximation to the preconditioned gradient, which is still effective~\cite{zhang-accelerating-2025-arxiv}.
On the other hand, the full construction and inversion of the preconditioner can be very beneficial at intermediate values of the bond dimension, because of the dramatic improvement it has on the required number of optimization steps.
In this work, we employ a hybrid approach: we explicitly construct the preconditioner at the beginning and then use it to precondition the linear systems for computing the preconditioned gradients in subsequent optimization steps. 
The explicit form of the preconditioner is reconstructed and updated only when the iterative solver fails to converge.

\subsection{Preparation of initial states} \label{subsec:preparation-of-initial-state}

In practice, we find that our optimization scheme often fails if it is directly applied to a random initial state. 
On the other hand, an ordinary L-BFGS optimization scheme can often decrease the energy very quickly at early stages of the optimization. 
Therefore, we first perform a number of ordinary optimization steps, which we refer to as preparation steps before we apply our improved optimization scheme. 
As no preconditioning is involved in the preparation steps, we can work in the gauge where all the $R$ matrices are diagonal and optimize the cMPS parameters directly. 
We provide a demonstration of the effect of the preparation steps in Appendix~\ref{sec:preparation-of-initial-state}.

\section{Benchmarks} \label{sec:benchmarks}

We benchmark the cMPS method in the two-component Lieb-Liniger model~\cite{lieb-exact-1963-a,quijandria-continuous-2014}.
The corresponding Hamiltonian is given by
\begin{align}
    H = \int \mathd x &\left[\sum_{\alpha=1,2} \frac{1}{2m_\alpha} \partial_x \psi_\alpha^\dagger(x) \partial_x \psi_\alpha(x) \right.\nonumber \\
    &+ \left.\sum_{\alpha, \beta=1,2} \frac{c_{\alpha \beta}}{2} \psi_\alpha^\dagger(x) \psi_\beta^\dagger(x) \psi_\beta(x) \psi_\alpha(x) \right], \label{eq:hamiltonian-two-component-lieb-liniger}
\end{align}
where $m_\alpha$ is the mass of the $\alpha$th bosonic species and $c_{\alpha \beta}$ is the interaction strength between the $\alpha$th and $\beta$th bosonic species.
Depending on the masses, interaction strengths, and particle densities of the two components, this model can exhibit a variety of physical phenomena~\cite{cazalilla-instabilities-2003}.

In the cMPS simulation, we cannot directly fix the particle density in the cMPS ansatz. 
Instead, we work in the grand canonical ensemble, where the particle density is controlled by an additional chemical potential term
\begin{equation}
    H_{\mathrm{chem}} = -\int \mathd x \left[\sum_{\alpha=1,2} \mu_\alpha \psi_\alpha^\dagger(x) \psi_\alpha(x) \right],
\end{equation}
where $\mu_\alpha$ is the chemical potential of the $\alpha$th bosonic species.

In this work, we restrict ourselves to the case where the Hamiltonian is ``symmetric'' with respect to the two components, i.e., $m_1 = m_2 = 1$, $c_{11} = c_{22} = c = 1$, and $\mu_1 = \mu_2 = \mu$.
For $c_{12} > c$, the system undergoes phase separation between the two components~\cite{quijandria-continuous-2015}. 
For $c_{12} < -c$, the system collapses due to the absence of a stable ground state. 
Hence, we focus exclusively on the parameter regime $-c < c_{12} < c$.

The effective Hamiltonian in the long-wavelength limit is then given by~\cite{cazalilla-instabilities-2003,cazalilla-bosonizing-2004}
\begin{align}
   H_{\mathrm{eff}} = & \int \mathd x \sum_{\alpha=1,2} \left[ \frac{\rho}{2}(\partial_x \phi_\alpha)^2 + \frac{\tilde{g}}{2\pi} (\partial_x \theta_\alpha)^2 \right] \nonumber \\
   & + \int \mathd x \left[ \frac{\tilde{g}_{12}}{\pi} (\partial_x \theta_1(x)) (\partial_x \theta_2(x))  \right. \nonumber \\
   & \hspace{4em} + \left. \frac{\tilde{g}_b}{2\pi} \cos \left(2 (\theta_1(x) - \theta_2(x))\right) \right],
\end{align}
where $\phi_\alpha$ and $\theta_\alpha$ are boson field operators corresponding to collective modes of phase and density fluctuations for bosons of type $\alpha$, respectively.
Since the Hamiltonian is symmetric with respect to the two components, the ground-state particle density for each type of boson is the same, i.e., $\rho_{1} = \rho_{2} = \rho$.
Here, $\tilde{g} = \tilde{g}_{11} = \tilde{g}_{22}$ is the effective intra-species interaction strength, and $\tilde{g}_{12}$ is the effective inter-species interaction strength.
The last cosine term originates from the inter-species backscattering process.

To diagonalize the effective Hamiltonian, one can introduce the following boson field operators~\cite{cazalilla-instabilities-2003}
\begin{equation}
    \phi_{\pm} = (\phi_1 \pm \phi_2) / \sqrt{2}, \quad \theta_{\pm} = (\theta_1 \pm \theta_2) / \sqrt{2},
\end{equation}
and the effective Hamiltonian becomes 
\begin{align}
    H_{\mathrm{eff}} = & \int \mathd x \left[ \frac{\rho}{2}(\partial_x \phi_+)^2 + \frac{\tilde{g} + \tilde{g}_{12}}{2\pi} (\partial_x \theta_+)^2 \right] \nonumber \\
    & + \int \mathd x \left[ \frac{\rho}{2}(\partial_x \phi_-)^2 + \frac{\tilde{g} - \tilde{g}_{12}}{2\pi} (\partial_x \theta_-)^2 + \frac{\tilde{g}_b}{2\pi} \cos \left(\sqrt{8} \theta_-\right) \right], 
\end{align}
where the effective Hamiltonian is decoupled into two sectors, which are usually referred to as the charge sector (denoted by subscript $+$) and the spin sector (denoted by subscript $-$).

We benchmark our cMPS method in this model from two (not necessarily disjoint) perspectives. 
First, in Sec.~\ref{sec:two-component-luttinger-liquid-theory}, we focus on the case where the system is far from the Tonks-Girardeau (TG) limit, such that the cosine term $\cos(\sqrt{8} \theta_-)$ is irrelevant in the long-wavelength limit~\cite{cazalilla-instabilities-2003}. 
In this case, the system is gapless in both charge and spin sectors, and is described by two decoupled Luttinger liquids for all values of $-c< c_{12} < c$.
We employ the cMPS method to study the critical properties of the system by computing the entanglement entropies, Luttinger parameters, sound velocities, and correlation functions.
Second, in Sec.~\ref{sec:attractive-inter-species-interactions}, we focus on the case of attractive inter-species interactions ($-c < c_{12} < 0$). 
In this regime, even with the chemical potential set to zero, the ground state exhibits a finite particle density, which is a purely quantum-mechanical effect, as a straightforward mean-field calculation would predict the vacuum state to be the ground state.
We then compute the ground-state particle density using our cMPS method and compare our results with those obtained from analytical calculations. 

\subsection{Critical properties in the two-component Luttinger liquid regime} \label{sec:two-component-luttinger-liquid-theory}

To keep the system away from the TG limit, we set the chemical potential to be $\mu = 2.0$, and the ground-state density is always large enough so that the dimensionless interaction strength $\gamma = c / \rho$ is sufficiently small.
In this case, the system is gapless in both charge and spin sectors, and is described by two decoupled Luttinger liquids for all values of $-c< c_{12} < c$.
In each sector, using the standard convention of the Luttinger liquid theory, the effective Hamiltonian can be rewritten as 
\begin{equation}
    H_{\mathrm{eff}, \pm} = \frac{v_\pm}{2\pi} \int \mathd x \left[ K_\pm (\partial_x \phi_\pm)^2 + \frac{1}{K_\pm} (\partial_x \theta_\pm)^2 \right],
\end{equation}
where $v_\pm$ is the sound velocity and $K_\pm$ is the Luttinger parameter.
One can easily verify that $v_\pm K_\pm = \pi \rho$ and $v_\pm / K_\pm = \tilde{g} \pm \tilde{g}_{12}$. 

\begin{figure}[!htb]
    \centering
    \resizebox{0.5\columnwidth}{!}{\includegraphics{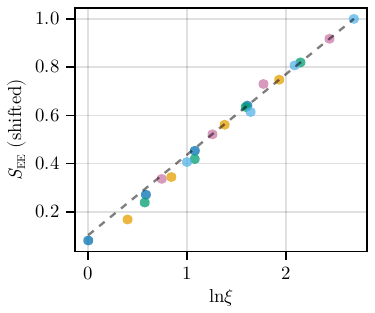}}
    \caption{Half-infinite entanglement entropy in the cMPS ground state approximation, plotted with respect to the logarithm of the cMPS correlation length $\xi$ (as naturally obtained for different bond dimensions), for $c_{12} = -0.6, -0.3, 0.0, 0.3, 0.6$ (shown in different colors).
    The bond dimensions of the cMPS data are $\chi = 8, 16, 32, 64$.
    The entanglement entropy data for different values of $c_{12}$ are vertically shifted by a constant so that the data points are collapsed onto a single line. 
    The dashed line represents the scaling $S_{\mathrm{EE}} = (C/6) \log(\xi) + S_0$, where $C=2$ is the central charge and $S_0$ is a constant.
    }
    \label{fig:entanglement}
\end{figure}

As a first check, we compute the entanglement scaling of the ground state using cMPS. 
The decoupled two-component Luttinger liquid theory has central charge $C=2$, and the half-infinite entanglement entropy should scale as $S_{\mathrm{EE}} = (C/6) \log \xi + S_0$, where $S_0$ is a constant~\cite{pollmann-theory-2009} and the finite correlation length $\xi$ is induced by the finite bond dimension (and thus entanglement) in the cMPS ansatz.
The results are shown in Fig.~\ref{fig:entanglement}, showing the entanglement entropies for cMPS with different bond dimensions and for different values of $c_{12}$.
By vertically shifting the data points from different values of $c_{12}$ by a constant, we can collapse the data points onto a single line, which confirms the linear scaling relation and central charge $C=2$.

Next, we compute the Luttinger parameter and the sound velocity from the stiffness, which is related to the second derivative of the energy density $\varepsilon$ with respect to the particle density $\rho_\pm$ in each sector. More specifically, we have~\cite{quijandria-continuous-2014}
\begin{equation}
   v_\pm = \sqrt{\rho \left(\frac{\partial^2 \varepsilon}{\partial \rho_\pm^2}\right)}, \; K_\pm = \pi \sqrt{\rho \left(\frac{\partial^2 \varepsilon}{\partial \rho_\pm^2}\right)^{-1} },
   \label{eq:stiffness}
\end{equation}
More details about the computation of the stiffness can be found in Appendix~\ref{sec:stiffness}.

\begin{figure}[!htb]
    \centering
    \resizebox{\columnwidth}{!}{\includegraphics{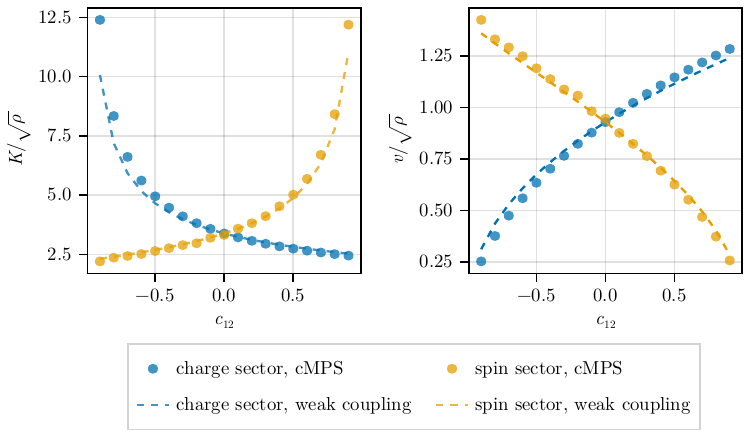}}
    \caption{The cMPS prediction for the Luttinger parameter $K_\pm$ and the sound velocity $v_\pm$ in the charge and spin sectors, respectively. 
    All the results are obtained with bond dimension $\chi = 32$. 
    The dashed lines represent the analytical results obtained from the weak-coupling expansion results \eqref{eq:weak-coupling-approximation-sound-velocity} and \eqref{eq:weak-coupling-approximation-luttinger-parameter}.
    }
    \label{fig:luttinger_parameter}
\end{figure}

Although there are no exact analytical results for $v_\pm$ and $K_\pm$, we can compare our numerical results with those obtained under certain approximations.
First, when the inter-species interaction strength is zero, the two boson species are independent of each other. 
If the dimensionless interaction strength $\gamma = c / \rho$ is small, within the weak-coupling expansion, the effective interaction strength in each sector can be approximated as $\tilde{g} = (c /\pi) (1 - \sqrt{\gamma} / 2\pi)$. 
Second, when $|c_{12}/c|$ is small, one can assume that $\tilde{g}$ remains unchanged compared to the case of $c_{12} = 0$, and $\tilde{g}_{12} = (c_{12} / c) \tilde{g}$.
Along this line, the Luttinger parameter and the sound velocity can be approximated as 
\begin{align}
    v_\pm &\approx \sqrt{\rho (c \pm c_{12})  \left(1 - \frac{\sqrt{\gamma}}{2\pi}\right)} , \label{eq:weak-coupling-approximation-sound-velocity} \\
    K_\pm &\approx \pi \sqrt{\rho} \left/ \sqrt{(c \pm c_{12}) \left(1 - \frac{\sqrt{\gamma}}{2\pi}\right)} \right. . \label{eq:weak-coupling-approximation-luttinger-parameter}
\end{align}
The ground-state density and the dimensionless interaction strength can be determined numerically. 

\begin{figure}[!htb]
    \centering
    \resizebox{\columnwidth}{!}{\includegraphics{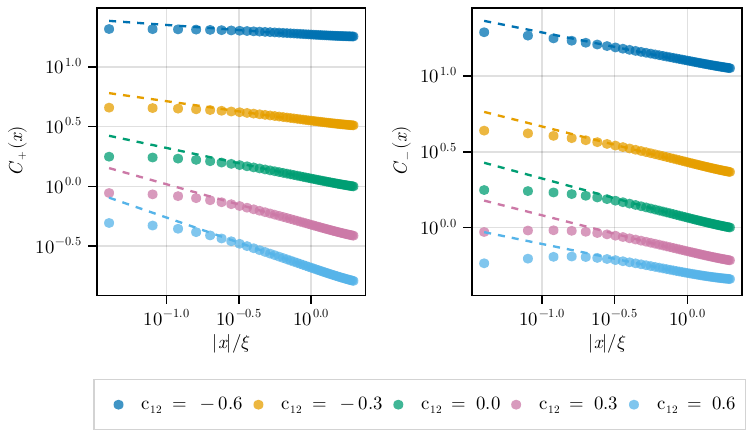}}
    \caption{The cMPS results for the correlation functions $C_+(x)$ and $C_-(x)$. 
    For each choice of $c_{12}$, the correlation functions are plotted with respect to the ratio between $x$ and the correlation length $\xi$ of the system. 
    The cMPS results are obtained with bond dimension $\chi=32$.
    The dashed lines represent the scaling predicted by the Luttinger liquid theory, where the values of $K_+$ and $K_-$ are taken from the numerical results obtained previously.
    }
    \label{fig:correlation_functions}
\end{figure}

The Luttinger parameters $K_+$ and $K_-$ determine the long-distance scaling behavior of correlation functions. 
More specifically, we consider the following two correlation functions 
\begin{align}
C_+(x) &= \left\langle \psi^\dagger_1(x) \psi_2^\dagger(x) \psi_2(0) \psi_1(0) \right\rangle, \\
C_-(x) &= \left\langle \psi^\dagger_1(x) \psi_2(x) \psi_2^\dagger(0) \psi_1(0) \right\rangle.
\end{align}
In the long-distance limit, these correlation functions exhibit power-law decay~\cite{cazalilla-instabilities-2003,cazalilla-bosonizing-2004},
\begin{equation}
C_+(x) \sim x^{-1/K_+}, \quad C_-(x) \sim x^{-1/K_-}.
\end{equation}
In Fig.~\ref{fig:correlation_functions}, we plot the cMPS results for $C_+(x)$ and $C_-(x)$ alongside the predicted scaling behaviors $x^{-1/K_+}$ and $x^{-1/K_-}$, where the values of $K_\pm$ are taken from the numerical results in Fig.~\ref{fig:luttinger_parameter}. 
The correlation functions obtained from the cMPS data agree very well with the expected power-law scaling in the long-distance regime.

\subsection{Ground-state density in the attractive inter-species regime} \label{sec:attractive-inter-species-interactions}

In the two-component Lieb-Liniger model \eqref{eq:hamiltonian-two-component-lieb-liniger},
a particularly interesting regime is the one with attractive inter-species interactions ($-c < c_{12} < 0$), where the particle density of the ground state is finite even when the chemical potential $\mu$ is zero.

In this regime, one can consider two different limits. 
The first limit is $|c_{12}/c| \ll 1$, where the system can be described by a TG gas of boson dimers~\cite{girardeau-relationship-1960,cazalilla-instabilities-2003,pricoupenko-dimerdimer-2018}.
In this case, the bosons of different species form boson dimers.
Each boson dimer has a length scale $\xi_{\mathrm{d}} \sim 1/|c_{12}|$ and contributes a negative binding energy $E_{\mathrm{d}} = -c_{12}^2 / 4$. 
In the TG limit, i.e., when the dimensionless interaction strength between dimers satisfies $\gamma_{\mathrm{d}} \gg 1$, the boson dimers can be mapped to non-interacting fermions~\cite{girardeau-relationship-1960}.
Along this line, the energy density of the system is given by~\cite{jiang-understanding-2015}
\begin{equation}
    \varepsilon_{\mathrm{TG}} = -E_{\mathrm{d}} \rho_{\mathrm{d}} + \frac{\pi^2}{12}\rho_{\mathrm{d}}^3,
\end{equation}
where $\rho_{\mathrm{d}} = \rho$ is the density of the boson dimers.

The other limit of interest is the limit of $1 - |c_{12}/c| \ll 1$, where the system is close to the collapse point $c_{12} = -c$.
In this limit, the ground-state density becomes very large, which indicates that the dimensionless interaction strengths $\gamma = c / \rho$ and $\gamma_{12} = c_{12} / \rho$ are both very small. In this weak-coupling regime, the system can be studied within the framework of Bogoliubov theory~\cite{bogoliubov-theory-1947,pitaevskii-bose-einstein-2016}. 
The energy density of the system is given by~\cite{petrov-ultradilute-2016,parisi-liquid-2019,parisi-quantum-2020}
\begin{equation}
    \varepsilon_{\mathrm{B}} = (c - |c_{12}|) \rho^2 - \frac{2\rho^{3/2}}{3\pi}\left[(c + |c_{12}|)^{3/2} + (c - |c_{12}|)^{3/2}\right],
\end{equation}
where the first term is the mean-field energy density and the second term is the beyond-mean-field Lee-Huang-Yang (LHY) correction~\cite{lee-eigenvalues-1957}.
This parameter regime has attracted a lot of interest recently, essentially due to the fact that the mean-field energy and the LHY correction have opposite signs and exhibit different scalings with the particle density.
This gives rise to a spinodal point in the energy function and indicates that, when the particle density is below the spinodal point, the uniform ground state becomes unstable and collapses to a self-bounded quantum droplet state even when no confining potential is applied~\cite{petrov-quantum-2015,petrov-ultradilute-2016,pricoupenko-dimerdimer-2018,parisi-liquid-2019,parisi-quantum-2020,englezos-multicomponent-2025}.

In the cMPS simulation, we always assume a uniform ground state, and compute it in the grand canonical ensemble, with the chemical potential $\mu$ as the control parameter.
More specifically, we consider the case where the chemical potential is set to $\mu = 0$, the ground-state density is finite, which is a purely quantum mechanical effect, as the mean-field treatment predicts an empty ground state. 
In the TG limit and the weak-coupling limit, the ground-state densities are given by
\begin{equation}
    \rho_{\mathrm{TG}} = \frac{|c_{12}|}{\pi}, \quad
    \rho_{\mathrm{B}} = \left(\frac{(c - |c_{12}|)^{3/2} + (c + |c_{12}|)^{3/2}}{2\pi (c - |c_{12}|)}\right)^2.
\end{equation}
One can \emph{a posteriori} verify that these results are consistent with the conditions for the TG limit and the weak-coupling limit, respectively. 

\begin{figure}[!htb]
    \centering
    \resizebox{\columnwidth}{!}{\includegraphics{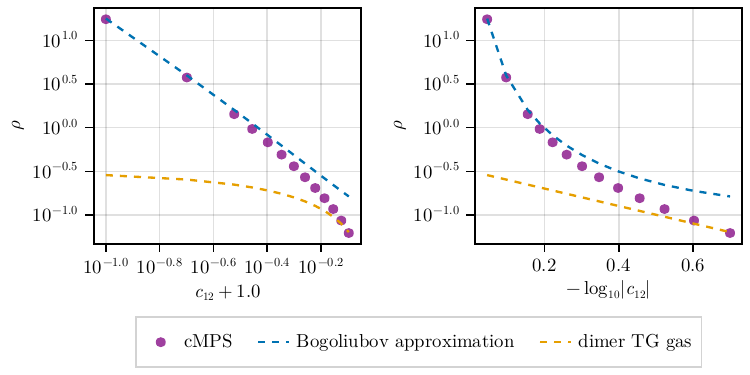}}
    \caption{The ground-state particle density $\rho$ as a function of the inter-species interaction strength $c_{12}$.
    We plot the results with respect to different functions of $c_{12}$, in regard to the TG limit and the weak-coupling limit, respectively.
    The cMPS results are obtained with bond dimension $\chi=32$.
    The dashed lines represent the analytical predictions in the two different limits. 
    }
    \label{fig:particle_density_inter_attractive_regime}
\end{figure}

In Fig.~\ref{fig:particle_density_inter_attractive_regime}, we compare the cMPS results with the analytical predictions in the two different limits.
As expected, the cMPS results agree well with the analytical predictions. 

\section{Summary and Outlook} \label{sec:summary-and-outlook}

To summarize, in this work, we have developed a new optimization scheme for the bosonic multi-component cMPS method, which enables us to perform cMPS simulations for multi-component boson systems with substantially larger bond dimensions than previous works, giving rise to more accurate numerical results.
We have benchmarked our approach in the two-component Lieb-Liniger model, and obtained numerical results that agree well with analytical predictions. 

There are several interesting future directions to explore with our multi-component cMPS method. In the two-component Luttinger liquid regime of the two-component Lieb-Liniger model, we have obtained Luttinger parameters by computing the stiffness. 
Recently, in the context of single-component Luttinger liquid systems, it has been shown that the Luttinger parameters can be directly obtained by computing the overlap between the ground state and the crosscap state~\cite{tan-extracting-2025}. 
Our multi-component cMPS method can be easily applied to investigate the crosscap overlap in cases of multi-component Luttinger liquids.

In our simulations of the two-component Lieb-Liniger model, we have restricted ourselves to the case where the Hamiltonian is ``symmetric'' with respect to the two components. 
It would be interesting to further study the properties of this model in other parameter regimes using the multi-component cMPS method.
Our approach can also be combined with the excitation ansatz to study the low-energy excitations in these systems.
Furthermore, it would also be interesting to explore whether the multi-component cMPS ansatz is sufficiently expressive to study early-time dynamical effects using the time-dependent variational principle.

More generally, it would be interesting to further extend our work to other quantum mixture systems, such as boson-fermion and fermion-fermion mixture systems. The constraints that need to be satisfied by the cMPS matrices are different in those cases, so that a suitable generic parametrization needs to be constructed. Finally, in order to make more direct contact with ongoing experiments, it would be useful to combine our approach with the techniques from Ref.~\cite{tuybens-variational-2022} in order to study inhomogeneous systems.

\section*{Acknowledgment}
We thank Karan Tiwana and Antoine Tilloy for valuable discussions. 
During our work, we became aware of their independent work on multi-field relativistic cMPS, which has been recently published as Ref.~\cite{tiwana-multi-field-2025-arxiv}. 
Their work is closely related to ours, but differs in that they focus on relativistic cMPS and employ a different strategy for treating the regularization condition.
We also thank Maarten Van Damme, Sophie Mutzel, and Frank Verstraete for helpful discussions.
W.~T. is grateful to Yueshui Zhang and Hong-Hao Tu for helpful discussions. 
W.~T. is funded by the Research Foundation Flanders postdoctoral Fellowship 12AA225N. B.~T. is supported as an `FWO-aspirant' under contract number FWO17/ASP/199. J.~H. receives funding from the European Research Council (ERC) under the European Union's Horizon 2020 program (grant agreement No.~101125822).

\paragraph*{Data availability}
The data used to produce the figures in this paper are available at \cite{tang_2025_18109141}.
\paragraph*{Code availability}
The code, used to calculate this multiple species cMPS optimization, will soon be available in the open-source Julia package CMPSKit.jl \footnote{\url{http://github.com/Jutho/CMPSKit.jl} (2021)}.

\appendix

\section{Imposing regularity conditions via penalty terms} \label{sec:penalty-method}

To remove the divergence in the kinetic energy \eqref{eq:kinetic-energy-divergence}, we can replace the divergent factor $\delta(0)$ with a finite cutoff $\Lambda$, and perform extrapolation to recover the correct result in the $\Lambda \rightarrow \infty$ limit. 
In practice, this is equivalent to removing the second term in Eq.~\eqref{eq:kinetic-energy-divergence} without introducing any constraints on the $R_j$ matrices, and then adding a penalty term $\hatH_{\mathrm{penalty}}(\Lambda)$ to the Hamiltonian.
This penalty term is given by
\begin{equation}
    \hatH_{\mathrm{penalty}}(\Lambda) = \Lambda \lim_{\epsilon \to 0} \int_{-\infty}^{+\infty} \mathd x O_{jk}(x,\epsilon)^\dagger O_{jk}(x, \epsilon).
\end{equation}
with
\begin{equation}
O_{jk}(x,\epsilon) = \psi_j(x) \psi_k(x+\epsilon) - \psi_k(x) \psi_j(x+\epsilon).
\end{equation}
Here, we have to evaluate the expectation value of this term with respect to the cMPS before taking the limit $\epsilon \to 0$. We refer to this regularization strategy as the penalty method.

This approach seems to offer several immediate advantages over the explicit parametrization approach. 
Firstly, the penalty method avoids imposing explicit constraints on the $R$ matrices, which makes it much easier to implement in practice.  
Secondly, the penalty method can be interpreted as a softened and weighted version of the regularity condition in Eq.~\eqref{eq:regularity-condition}, as it takes the environment of the $R$ matrices into account.
Within the broader context of tensor networks, this might seem as a very natural generalization.

However, during the penalty method simulation, one eventually needs to extrapolate the result to the $\Lambda \rightarrow \infty$ limit, and inevitably needs to perform simulations with large $\Lambda$ values. 
When $\Lambda \gg 1$, the penalty term makes the energy function very ill-conditioned, resulting in a very slow optimization process.
Indeed, in this limit, the regularity conditions need again to be satisfied more and more accurately, and a general gradient optimization method has a hard time finding and staying on this lower-dimensional submanifold of commuting $R_j$ matrices.
This is in sharp contrast to ordinary cMPS simulations, where the ill-conditioning of the energy landscape mostly originates from the geometry of the cMPS ansatz itself.
For this reason, the typical choice of tangent-space metric preconditioner is not effective in these simulations. 

\section{Regularization parameter in the preconditioner} \label{sec:regularization-parameter-in-preconditioner}

The regularization parameter $\lambda$ in the preconditioner \eqref{eq:preconditioner-tangent-vectors-left-canonical} can have a large impact on the performance of the optimization in certain parameter regimes. 
In practice, it has been observed that it is often beneficial to start with a relatively large regularization parameter at the beginning of the optimization, and gradually decrease it during the optimization process. However, a truly satisfactory explanation of why this is the case, and why the pure tangent-space metric preconditioner ($\lambda=0$) can actually slow down the convergence in the initial part of the optimization, is currently unknown~\cite{hauru-riemannian-2021,tang-kac-2023,zhang-accelerating-2025-arxiv}. 

Nevertheless, in the context of cMPS simulations, there is at least one guiding principle that can be used when determining the value of $\lambda$, which is that the optimization should be covariant under coordinate transformations of the problem.
More specifically, we can consider the coordinate transformation 
\begin{equation}
x \rightarrow \tilde{x} = (1/\kappa) x, \label{eq:coordinate-transformation}
\end{equation}
where $\kappa$ is a scaling factor.
Under this coordinate transformation, the relevant Hamiltonian parameters (interaction strength $c$ and chemical potential $\mu$) and observables (particle density $\rho$ and energy density $\varepsilon$) transform accordingly as 
\begin{equation}
    c \rightarrow \kappa c, \quad \mu \rightarrow \kappa^2 \mu , \quad \rho \rightarrow \kappa \rho , \quad \varepsilon \rightarrow \kappa^3  \varepsilon \label{eq:hamiltonian-parameter-transformation},
\end{equation}
and the cMPS parameters transform as 
\begin{equation}
    Q \rightarrow \kappa Q, \quad M \rightarrow M, \quad D_j \rightarrow \sqrt{\kappa} D_j. \label{eq:cMPS-parameter-transformation}
\end{equation}
As a result, the tangent vector parameters $\delta X$ and $\delta D$ should also transform accordingly as 
\begin{equation}
    \delta X \rightarrow \kappa^3 \delta X, \quad \delta D \rightarrow \kappa^2 \sqrt{\kappa} \, \delta D. \label{eq:tangent-vector-transformation}
\end{equation}

One can verify that the original tangent-space metric preconditioner \eqref{eq:preconditioner-tangent-vectors-left-canonical} is covariant under the coordinate transformation \eqref{eq:coordinate-transformation}, i.e., the search direction determined by the preconditioner \eqref{eq:preconditioner-tangent-vectors-left-canonical} is invariant under coordinate transformations up to scaling factors in Eq.~\eqref{eq:cMPS-parameter-transformation}.
To see this, one can notice that the tangent-space metric $\mathcal{N}$ has the following structure [cf. Eq.~\eqref{eq:preconditioner-tangent-vectors-left-canonical}]
\begin{equation}
    \mathcal{N} = \begin{pmatrix}
        \mathcal{N}_{\delta X, \delta X} & \mathcal{N}_{\delta X, \delta D} \\
        \mathcal{N}_{\delta D, \delta X} & \mathcal{N}_{\delta D, \delta D}
    \end{pmatrix}. \label{eq:tangent-space-metric-structure}
\end{equation}
Here, we have defined $\mathcal{N}_{\delta X, \delta X}$ as the matrix elements of $\mathcal{N}$ that are associated with the tangent vector overlaps where $\delta X^a$ and $\delta X^b$ are nonzero, but $\delta D_j^a$ and $\delta D_j^b$ are zero [cf. Eq.~\eqref{eq:preconditioner-tangent-vectors-left-canonical}].
The matrix elements $\mathcal{N}_{\delta X, \delta D}$, $\mathcal{N}_{\delta D, \delta X}$, and $\mathcal{N}_{\delta D, \delta D}$ are defined similarly.
Under the coordinate transformation \eqref{eq:coordinate-transformation}, the matrix elements of $\mathcal{N}$ transform as 
\begin{align}
    \mathcal{N}_{\delta X, \delta X} \rightarrow \kappa \mathcal{N}_{\delta X, \delta X}, \quad & \mathcal{N}_{\delta X, \delta D} \rightarrow \sqrt{\kappa} \mathcal{N}_{\delta X, \delta D}, \quad  \label{eq:tangent-space-metric-transformation1} \\
    \mathcal{N}_{\delta D, \delta X} \rightarrow \sqrt{\kappa} \mathcal{N}_{\delta D, \delta X}, \quad & \mathcal{N}_{\delta D, \delta D} \rightarrow \mathcal{N}_{\delta D, \delta D}. \label{eq:tangent-space-metric-transformation2}
\end{align}
Combining Eqs.~\eqref{eq:tangent-vector-transformation}, \eqref{eq:tangent-space-metric-structure}, \eqref{eq:tangent-space-metric-transformation1}, and \eqref{eq:tangent-space-metric-transformation2}, one can verify that the preconditioned gradient is covariant under the coordinate transformation \eqref{eq:coordinate-transformation}.

However, a naive choice of the regularization parameter $\lambda$ can easily break this covariance. 
This could lead to suboptimal performance of the optimization in certain parameter regimes, especially when the ground-state density is very large or very small [cf. Eq.~\eqref{eq:hamiltonian-parameter-transformation}].
To restore the covariance, one should choose the regularization term in a way such that the regularized metric still satisfies the transformation rules \eqref{eq:tangent-space-metric-transformation1} and \eqref{eq:tangent-space-metric-transformation2}.

In our simulations, we choose different regularization parameters for the blocks $\mathcal{N}_{\delta X, \delta X}$ and $\mathcal{N}_{\delta D, \delta D}$ separately, i.e., $\mathcal{N}_{\delta X, \delta X} \rightarrow \mathcal{N}_{\delta X, \delta X} + \lambda_{\delta X, \delta X} \mathcal{I}_{\delta X, \delta X}$ and $\mathcal{N}_{\delta D, \delta D} \rightarrow \mathcal{N}_{\delta D, \delta D} + \lambda_{\delta D, \delta D} \mathcal{I}_{\delta D, \delta D}$, where $\lambda_{\delta X, \delta X}$ and $\lambda_{\delta D, \delta D}$ are regularization parameters, and $\mathcal{I}_{\delta X, \delta X}$ and $\mathcal{I}_{\delta D, \delta D}$ are identity matrices.
First, we introduce a factor $\lambda_\varepsilon = |\varepsilon|^{1/3}$, which scales as $\lambda_\varepsilon \rightarrow \kappa \lambda_\varepsilon$ under the coordinate transformation \eqref{eq:coordinate-transformation}~\footnote{This choice does not work if the ground-state energy density is zero, which does not happen in our simulations. Alternatively, one can also use different factors according to other quantities, e.g. the particle density $\rho$. We have not explored and compared these options in full detail, as the current choice works well in most of our simulations. }.
We then use it to rescale the gradient as $\delta X^g \rightarrow \lambda_\varepsilon^{-3} \delta X^g$ and $\delta D^g \rightarrow \lambda_\varepsilon^{-5/2} \delta D^g$.
The regularization parameter $\lambda_{\delta D, \delta D}$ is then chosen as the square of the norm of the rescaled gradient, i.e., $\lambda_{\delta D, \delta D} = \lambda_\varepsilon^{-6} |\delta X^g|^2 + \lambda_\varepsilon^{-5} \sum_j |\delta D^g_j|^2$, while the regularization parameter $\lambda_{\delta X, \delta X}$ is set to $\lambda_{\delta X, \delta X} = \lambda_\varepsilon \lambda_{\delta D, \delta D}$.

\section{Preparation of the initial state} \label{sec:preparation-of-initial-state}

\begin{figure}[!htb]
    \centering
    \resizebox{0.9\columnwidth}{!}{\includegraphics{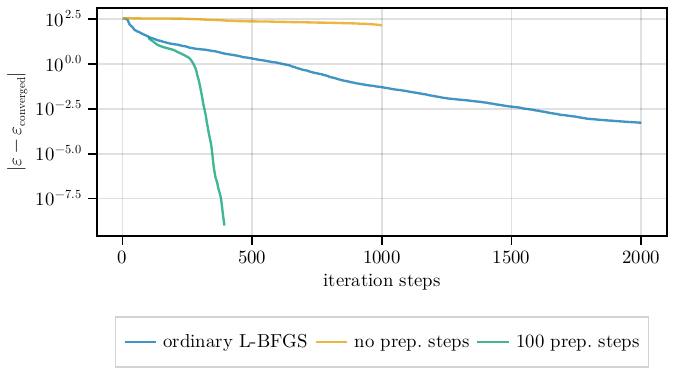}}
    \caption{A comparison of the optimization performance of our optimization scheme with and without preparation steps. 
    As a comparison, we also plot the energy curve of the ordinary L-BFGS optimization scheme in the diagonal gauge.
    The bond dimension of the cMPS is $\chi=8$, and the Hamiltonian parameters are $c = 10.0$, $c_{12} = -7.0$, and $\mu = 0.0$.}
    \label{fig:preparation_steps}
\end{figure}

Here, we provide a benchmark of the optimization performance of our optimization scheme with and without preparation steps.
As shown in Fig.~\ref{fig:preparation_steps}, we find that, if directly applied to a random initial state, the performance of our optimization scheme is actually much worse than the ordinary L-BFGS optimization scheme in the diagonal gauge.
However, if we first perform a number of ordinary optimization steps to prepare the initial state, the performance of our optimization scheme is much improved and can quickly converge. 

The number of preparation steps required is unclear \emph{a priori}, and it depends on the Hamiltonian, the initial state, and the bond dimension of the cMPS. 
Nevertheless, in practice, we find that, in most cases, the number of preparation steps required only increases moderately with the bond dimension of the cMPS, as the ordinary L-BFGS optimization typically works very well at early stages of the optimization.

\section{Computation of the stiffness} \label{sec:stiffness}

This final appendix provides a more detailed discussion of the computation of the stiffness \eqref{eq:stiffness} in the two-component Lieb-Liniger model.
We compute the second derivative of the energy density with respect to the particle density by numerical differentiation. 

For the stiffness in the charge sector, we change the particle density by adding small perturbations $\delta \mu$ to the chemical potentials $\mu_1$ and $\mu_2$ as $\mu_{1,2} \rightarrow \mu_{1,2} + \delta \mu$. 
As the perturbed Hamiltonian is still ``symmetric'' with respect to the two components, the density in the spin sector remains zero, i.e., $\rho_{-} = 0$.
The stiffness is then directly computed by numerical differentiation of the energy density with respect to the charge sector density $\rho_{+}$. 

For the stiffness in the spin sector, we perturb the chemical potentials $\mu_{1,2}$ in opposite directions, i.e., $\mu_1 \rightarrow \mu_1 + \delta \mu$ and $\mu_2 \rightarrow \mu_2 - \delta \mu$.
This induces nonzero density values in the spin sector $\rho_{-}$, but also slightly changes the density in the charge sector $\rho_{+}$. 
To ensure that the density in the charge sector $\rho_{+}$ remains unchanged, we use the coordinate transformation \eqref{eq:coordinate-transformation} to rescale the densities, so that $\rho_{+}$ remains the same as in the unperturbed system. 
We then use the scaled particle densities and energy density to compute the second-order derivative of the energy density with respect to the spin sector density $\rho_{-}$. 

In our simulations, the second-order derivatives are computed from 7 data points, where $\delta \mu$ takes values $0, \pm 0.005, \pm 0.01, \pm 0.015$.

\bibliography{multiboson}

\end{document}